\documentclass[12pt]{article}

\usepackage{amsmath,amssymb,amscd,amsbsy,amsgen,amsopn,amstext,
amsxtra}

\usepackage[dvips]{graphicx}

\begin{document}

\begin{flushright}
May, 2005
\end{flushright}

\vskip 0.5 truecm

\begin{center}
{\Large{\bf Geometric phases and hidden local
gauge symmetry}}
\end{center}
\vskip .5 truecm
\centerline{\bf  Kazuo Fujikawa }
\vskip .4 truecm
\centerline {\it Institute of Quantum Science, College of 
Science and Technology}
\centerline {\it Nihon University, Chiyoda-ku, Tokyo 101-8308, 
Japan}
\vskip 0.5 truecm

\makeatletter
\@addtoreset{equation}{section}
\def\theequation{\thesection.\arabic{equation}}
\makeatother

\begin{abstract}
The analysis of geometric phases
 associated with level crossing is reduced to the familiar 
diagonalization of the Hamiltonian in the second quantized 
formulation. 
A hidden local gauge symmetry, which is associated with the 
arbitrariness of the phase choice of a complete orthonormal 
basis set, becomes explicit in this formulation (in particular,
in the adiabatic approximation)
 and specifies physical observables. The choice of a basis set 
which specifies the coordinate in the functional space is 
arbitrary in the second quantization, and a sub-class of coordinate 
transformations, which keeps the form of the action invariant,
 is recognized as the gauge symmetry. 
We  discuss the implications of this hidden local gauge 
symmetry in detail by analyzing geometric phases for cyclic and 
noncyclic evolutions. 
 It is shown that the hidden local symmetry provides a basic 
concept alternative to the notion of holonomy to analyze 
geometric 
phases and that the analysis based on the hidden local 
gauge symmetry leads to results consistent with the general 
prescription  of Pancharatnam. We however note 
an important difference between the geometric phases for cyclic 
and noncyclic evolutions. We also explain  a basic 
difference between our hidden local gauge symmetry and a gauge 
symmetry (or equivalence class) used by Aharonov and Anandan in 
their definition of generalized geometric phases. 
\end{abstract}


\section{Introduction}
The geometric phases have been mainly analyzed in the framework 
of first quantization by using the adiabatic 
approximation~\cite{berry}-\cite{bhandari}, though the processes
slightly away from  adiabaticity have been considered in 
\cite{berry2} and a definition of generalized phase, which does 
not explicitly refer to the adiabatic approximation,
 has been given in~\cite{aharonov}. 
Interesting mathematical ideas such as  parallel transport and 
holonomy are also introduced in the framework of 
adiabatic approximation~\cite{simon}. In the precise adiabatic limit, the phase becomes  non-dynamical and  geometric.
A generalization of geometric phases 
for noncylcic evolutions has also been proposed~\cite{samuel}. 
The old idea of Pancharatnam~\cite{pancharatnam, ramaseshan, 
berry3} plays an important role in this generalization. These
earlier works have been further elaborated by various authors,
for example, in Refs.~\cite{giavarini,aitchison, mukunda, pati1, pati2, garcia, mostafazadeh} and references therein.

It has been recently shown~\cite{fujikawa, fujikawa2} that a 
second 
quantized formulation provides a convenient framework for the analysis of geometric 
phases without assuming the adiabatic approximation. In this 
formulation, the analysis of geometric phases is reduced to a
diagonalization of the Hamiltonian, namely, the geometric phases
 become parts of the dynamical phases. See also 
Ref.~\cite{berry2} for a possible dynamical interpretation 
of geometric phases. One recovers the conventional geometric 
phases defined in the adiabatic approximation when one 
diagonalizes the Hamiltonian in a very specific limit.  If one 
diagonalizes the 
Hamiltonian in the other extreme limit, namely, in the 
infinitesimal neighborhood of level crossing for any fixed 
finite time interval $T$, the geometric phases become trivial 
and thus no monopole-like singularity. At the level crossing 
point, the conventional energy eigenvalues become degenerate 
but the degeneracy is lifted if one diagonalizes the geometric 
terms~\footnote{In passing, we note that the degeneracy analyzed
 in the geometric phases and the 
{\em non-level crossing theorem}~\cite{sakurai} have no direct 
connection. In the $2\times 2$ traceless hermitian matrix, for 
example, the latter theorem states that the nondegenerate 
diagonal eigenvalues do not become degenerate by simply varying 
the off-diagonal elements. In the analysis of geometric phases, 
the level crossing is defined by the point where all the 
matrix elements vanish. }. The topological 
interpretation~\cite{stone, berry} of 
geometric phases such as the topological proof of the 
Longuet-Higgins' phase-change rule~\cite{higgins}, for example,
thus fails in the practical Born-Oppenheimer approximation where 
$T$ is identified with the period of the slower system. For a 
fixed finite $T$, the phases cease to be purely geometric.

Interpreted as a dynamical phase, the geometric phase appears 
in any process, regardless of non-adiabatic or noncyclic 
evolutions. In the present paper, we discuss the implications of
 the hidden local gauge symmetry, which appears in the second 
quantized formulation as a result of the arbitrariness
of the phase choice of the complete orthonormal basis set.
This gauge symmetry originates in the fact that the choice of a
basis set which specifies the coordinate in the functional space
is arbitrary in the second quantization as long as the 
coordinate is not singular, and thus the sub-class of coordinate 
transformations which preserves the form of the action
is recognized as a gauge symmetry. This hidden local gauge 
symmetry is an exact symmetry of quantized theory, and its
essence in the analysis of geometric phases has been briefly described  in Ref.~\cite{fujikawa2}. We 
here discuss its full implications in the analysis of geometric 
phases including noncyclic evolutions in general.  
This hidden local symmetry specifies physical observables. It is
 shown that the hidden local 
gauge symmetry provides a basic concept alternative to the 
notions of parallel transport and holonomy to analyze geometric 
phases and that the consideration on the basis of the local 
symmery leads to results consistent with the general 
prescription of Pancharatnam. 
In the course of our analysis, we mention some of the related 
past works~\cite{giavarini,aitchison, mukunda, pati1, pati2, garcia, mostafazadeh} in
the framework of first quantization,
though the notion of the hidden local symmetry itself has not 
been stated in these works. We also compare in detail this 
hidden 
local gauge symmetry to a local gauge symmetry (or equivalence 
class) considered by Aharonov and Anandan~\cite{aharonov} and 
also by Samuel and Bhandari~\cite{samuel}, which 
changes the form of the Schr\"{o}dinger equation and thus not a 
symmetry of quantized  theory in the conventional sense. 

\section{Second quantized formulation and geometric phases}

We start with the generic (hermitian) Hamiltonian 
\begin{equation}
\hat{H}=\hat{H}(\hat{\vec{p}},\hat{\vec{x}},X(t))
\end{equation}
for a single particle theory in a slowly varying background 
variable $X(t)=(X_{1}(t),X_{2}(t),...)$.
The path integral for this theory for the time interval
$0\leq t\leq T$ in the second quantized 
formulation is given by 
\begin{eqnarray}
Z&=&\int{\cal D}\psi^{\star}{\cal D}\psi
\exp\{\frac{i}{\hbar}\int_{0}^{T}dtd^{3}x[
\psi^{\star}(t,\vec{x})i\hbar\frac{\partial}{\partial t}
\psi(t,\vec{x})\nonumber\\
&&-\psi^{\star}(t,\vec{x})
\hat{H}(\frac{\hbar}{i}\frac{\partial}{\partial\vec{x}},
\vec{x},X(t))\psi(t,\vec{x})] \}.
\end{eqnarray}
We then define a complete set of eigenfunctions
\begin{eqnarray}
&&\hat{H}(\frac{\hbar}{i}\frac{\partial}{\partial\vec{x}},
\vec{x},X(0))u_{n}(\vec{x},X(0))
=\lambda_{n}u_{n}(\vec{x},X(0)), \nonumber\\
&&\int d^{3}xu_{n}^{\star}(\vec{x},X(0))u_{m}(\vec{x},X(0))=
\delta_{nm},
\end{eqnarray}
and expand 
\begin{eqnarray}
\psi(t,\vec{x})=\sum_{n}a_{n}(t)u_{n}(\vec{x},X(0)).
\end{eqnarray}
We then have
\begin{eqnarray}
{\cal D}\psi^{\star}{\cal D}\psi=\prod_{n}{\cal D}a_{n}^{\star}
{\cal D}a_{n}
\end{eqnarray}
and the path integral is written as 
\begin{eqnarray}
Z&=&\int \prod_{n}{\cal D}a_{n}^{\star}
{\cal D}a_{n}
\exp\{\frac{i}{\hbar}\int_{0}^{T}dt[
\sum_{n}a_{n}^{\star}(t)i\hbar\frac{\partial}{\partial t}
a_{n}(t)\nonumber\\
&&-\sum_{n,m}a_{n}^{\star}(t)E_{nm}(X(t))a_{m}(t)] \}
\end{eqnarray}
where 
\begin{eqnarray}
E_{nm}(X(t))=\int d^{3}x u_{n}^{\star}(\vec{x},X(0))
\hat{H}(\frac{\hbar}{i}\frac{\partial}{\partial\vec{x}},
\vec{x},X(t))u_{m}(\vec{x},X(0)).
\end{eqnarray}

We next perform a unitary transformation
\begin{eqnarray}
a_{n}(t)=\sum_{m}U(X(t))_{nm}b_{m}(t)
\end{eqnarray}
where 
\begin{eqnarray}
U(X(t))_{nm}=\int d^{3}x u^{\star}_{n}(\vec{x},X(0))
v_{m}(\vec{x},X(t))
\end{eqnarray}
with the instantaneous eigenfunctions of the Hamiltonian
\begin{eqnarray}
&&\hat{H}(\frac{\hbar}{i}\frac{\partial}{\partial\vec{x}},
\vec{x},X(t))v_{n}(\vec{x},X(t))
={\cal E}_{n}(X(t))v_{n}(\vec{x},X(t)), \nonumber\\
&&\int d^{3}x v^{\star}_{n}(\vec{x},X(t))v_{m}(\vec{x},X(t))
=\delta_{n,m}.
\end{eqnarray}
We emphasize that $U(X(t))$ may be chosen to be a unit matrix 
both at $t=0$ and $t=T$ if $X(T)=X(0)$, and thus 
\begin{eqnarray}
\{a_{n}\}=\{b_{n}\}
\end{eqnarray}
both at $t=0$ and $t=T$. We take the time $T$ 
as a period of the slowly varying variable $X(t)$ in the analysis
of geometric phases, unless stated otherwise.
We call the phase choice 
$v_{n}(\vec{x},X(t))$ in (2.10) as a standard basis set, and 
the more general choice of phase will be discussed 
later in connection with the hidden local gauge symmetry.
We can thus re-write the path integral as 
\begin{eqnarray}
&&Z=\int \prod_{n}{\cal D}b_{n}^{\star}{\cal D}b_{n}
\exp\{\frac{i}{\hbar}\int_{0}^{T}dt[
\sum_{n}b_{n}^{\star}(t)i\hbar\frac{\partial}{\partial t}
b_{n}(t)\nonumber\\
&&+\sum_{n,m}b_{n}^{\star}(t)
\langle n|i\hbar\frac{\partial}{\partial t}|m\rangle
b_{m}(t)-\sum_{n}b_{n}^{\star}(t){\cal E}_{n}(X(t))b_{n}(t)] \}
\end{eqnarray}
where the second term in the action stands for the term
commonly referred to as Berry's phase\cite{berry} and its 
off-diagonal generalization. 
The second term in (2.12) is defined by
\begin{eqnarray} 
(U(X(t))^{\dagger}i\hbar\frac{\partial}{\partial t}U(X(t)))_{nm}
&=&\int d^{3}x v^{\star}_{n}(\vec{x},X(t))
i\hbar\frac{\partial}{\partial t}v_{m}(\vec{x},X(t))\nonumber\\
&\equiv& \langle n|i\hbar\frac{\partial}{\partial t}|m\rangle.
\end{eqnarray}
The path integral (2.12) is also derived directly by expanding $\psi(t,\vec{x})
=\sum_{n}b_{n}(t)v_{n}(\vec{x},X(t))$ in terms of  the 
instantaneous eigenfunctions in (2.10).  

In the operator formulation of the second quantized theory,
we thus obtain the effective Hamiltonian (depending on Bose or 
Fermi statistics)
\begin{eqnarray}
\hat{H}_{eff}(t)&=&\sum_{n}\hat{b}_{n}^{\dagger}(t)
{\cal E}_{n}(X(t))\hat{b}_{n}(t)\nonumber\\
&&-\sum_{n,m}\hat{b}_{n}^{\dagger}(t)
\langle n|i\hbar\frac{\partial}{\partial t}|m\rangle
\hat{b}_{m}(t)
\end{eqnarray}
with 
\begin{eqnarray}
[\hat{b}_{n}(t), \hat{b}^{\dagger}_{m}(t)]_{\mp}=\delta_{n,m}.
\end{eqnarray}
Note that these formulas (2.6), (2.12) and (2.14) are 
exact. See also Ref.~\cite{anandan} for a formula related to 
(2.14) in the first quantization. The use of the instantaneous 
eigenfunctions in (2.12) is
 a common feature shared with the adiabatic approximation. In 
our picture,
 all the information about geometric phases  is included in 
the effective Hamiltonian, and for this reason we use the 
terminology ``geometric terms'' for those general terms 
appearing in the Hamiltonian. The ``geometric phases'' are used
when these terms are interpreted as phase factors of a specific
state vector. The fact that the Berry's phase can be understood as a part of the Hamiltonian, i.e.,{\em dynamical}, has been noted in an adiabatic picture~\cite{berry2}.
Our formula does not assume the adiabatic approximation, and 
thus it gives a generalization.

When one defines the Schr\"{o}dinger picture by
\begin{eqnarray}
\Psi_{S}(t)&=&U(t)^{\dagger}\Psi_{H}(0),\nonumber\\
\hat{b}_{n}(0)&=&U(t)^{\dagger}\hat{b}_{n}(t)U(t),\nonumber\\
\hat{{\cal H}}_{eff}(t)&\equiv& 
U(t)^{\dagger}\hat{H}_{eff}(t)U(t)
\nonumber\\
&=&\sum_{n}\hat{b}_{n}^{\dagger}(0)
{\cal E}_{n}(X(t))\hat{b}_{n}(0)
-\sum_{n,m}\hat{b}_{n}^{\dagger}(0)
\langle n|i\hbar\frac{\partial}{\partial t}|m\rangle
\hat{b}_{m}(0),
\end{eqnarray}
where
\begin{eqnarray}
i\hbar\frac{\partial}{\partial t}U(t)= - \hat{H}_{eff}(t)U(t)
\end{eqnarray}
with $U(0)=1$ (and thus $i\hbar\frac{\partial}{\partial t}U(t)
= -U(t) \hat{{\cal H}}_{eff}(t)$),
the second quantization formula for the evolution operator 
gives rise to~\cite{fujikawa, fujikawa2}  
\begin{eqnarray}
&&\langle n|T^{\star}\exp\{-\frac{i}{\hbar}\int_{0}^{T}
\hat{{\cal H}}_{eff}(t)
dt\}|n\rangle\nonumber\\ 
&&=
\langle n(T)|T^{\star}\exp\{-\frac{i}{\hbar}\int_{0}^{T}
\hat{H}(\hat{\vec{p}}, \hat{\vec{x}},  
X(t))dt \}|n(0)\rangle \,
\end{eqnarray}
where $T^{\star}$ stands for the time ordering operation, and 
the state vectors in the second quantization  on the left-hand 
side are defined by 
\begin{eqnarray}
|n\rangle=\hat{b}_{n}^{\dagger}(0)|0\rangle,
\end{eqnarray} 
and the state vectors on the right-hand side  stand for the 
first quantized states defined by
\begin{eqnarray}
\hat{H}(\hat{\vec{p}}, \hat{\vec{x}},  
X(t))|n(t)\rangle
={\cal E}_{n}(X(t))|n(t)\rangle.
\end{eqnarray}
Both-hand sides of the above equality (2.18) are exact, but the 
difference is that the geometric terms, both of diagonal and 
off-diagonal, are explicit in the second quantized formulation 
on the left-hand side.

The relation (2.18) is generalized for the off-diagonal elements
 also by following the same procedure in~\cite{fujikawa, fujikawa2}
\begin{eqnarray}
&&\langle m|T^{\star}\exp\{-\frac{i}{\hbar}\int_{0}^{T}
\hat{{\cal H}}_{eff}(t)
dt\}|n\rangle\nonumber\\ 
&&=
\langle m(T)|T^{\star}\exp\{-\frac{i}{\hbar}\int_{0}^{T}
\hat{H}(\hat{\vec{p}}, \hat{\vec{x}},  
X(t))dt \}|n(0)\rangle .
\end{eqnarray}
By noting
\begin{eqnarray}
v_{m}(\vec{x};X(T))=\langle \vec{x}|m(T)\rangle,
\end{eqnarray}
we define
\begin{eqnarray}
\psi_{n}(\vec{x},T; X(T))&\equiv&\sum_{m}v_{m}(\vec{x};X(T))
\langle m|T^{\star}\exp\{-\frac{i}{\hbar}\int_{0}^{T}
\hat{{\cal H}}_{eff}(t)dt\}|n\rangle\nonumber\\
&=&\sum_{m}v_{m}(\vec{x};X(T))
\langle m(T)|T^{\star}\exp\{-\frac{i}{\hbar}\int_{0}^{T}
\hat{H}(\hat{\vec{p}}, \hat{\vec{x}},  
X(t))dt \}|n(0)\rangle\nonumber\\
&=&\langle \vec{x}|T^{\star}\exp\{-\frac{i}{\hbar}\int_{0}^{T}
\hat{H}(\hat{\vec{p}}, \hat{\vec{x}},  
X(t))dt \}|n(0)\rangle.
\end{eqnarray}
This $\psi_{n}(\vec{x},T; X(T))$ satisfies the 
equation
\begin{eqnarray}
i\hbar\frac{\partial}{\partial T}\psi_{n}(\vec{x},T; X(T))
&=&i\hbar\frac{\partial}{\partial T}
\langle \vec{x}|T^{\star}\exp\{-\frac{i}{\hbar}\int_{0}^{T}
\hat{H}(\hat{\vec{p}}, \hat{\vec{x}},  X(t))dt \}|n(0)\rangle
\nonumber\\
&=&\langle \vec{x}|\hat{H}(\hat{\vec{p}}, \hat{\vec{x}},  X(T))
T^{\star}\exp\{-\frac{i}{\hbar}\int_{0}^{T}
\hat{H}(\hat{\vec{p}}, \hat{\vec{x}},  X(t))dt \}|n(0)\rangle
\nonumber\\
&=&\hat{H}(\frac{\hbar}{i}\frac{\partial}{\partial\vec{x}}, 
\vec{x},  X(T))\langle \vec{x}|T^{\star}
\exp\{-\frac{i}{\hbar}\int_{0}^{T}
\hat{H}(\hat{\vec{p}}, \hat{\vec{x}},  X(t))dt \}|n(0)\rangle
\nonumber\\
&=&\hat{H}(\frac{\hbar}{i}\frac{\partial}{\partial\vec{x}}, 
\vec{x},  X(T))\psi_{n}(\vec{x},T; X(T))
\end{eqnarray}
with the initial condition
\begin{eqnarray}
\psi_{n}(\vec{x},0; X(0))=v_{n}(\vec{x}; X(0)).
\end{eqnarray}
The amplitude $\psi_{n}(\vec{x},T; X(T))$ thus corresponds to 
the probability amplitude we deal with in the analysis of 
geometric phases. 
This $\psi_{n}(\vec{x},T; X(T))$ is also written as 
\begin{eqnarray}
\psi_{n}(\vec{x},T; X(T))=
\langle 0|\hat{\psi}(T,\vec{x})\hat{b}^{\dagger}_{n}(0)|0\rangle
\end{eqnarray}
by noting
\begin{eqnarray}
\hat{b}_{m}(T)&=&\left(T^{\star}\exp\{-\frac{i}{\hbar}
\int_{0}^{T}
\hat{{\cal H}}_{eff}(t)dt\}\right)^{\dagger}\hat{b}_{m}(0)
\left(T^{\star}\exp\{-\frac{i}{\hbar}\int_{0}^{T}
\hat{{\cal H}}_{eff}(t)dt\}\right),\nonumber\\
\hat{\psi}(T,\vec{x})&=&\sum_{m}v_{m}(\vec{x};X(T))\hat{b}_{m}(T)
\end{eqnarray}
and thus 
\begin{eqnarray}
\langle 0|\hat{\psi}(T,\vec{x})\hat{b}^{\dagger}_{n}(0)
|0\rangle&=&
\sum_{m}v_{m}(\vec{x};X(T))\langle 0|\hat{b}_{m}(T)
\hat{b}^{\dagger}_{n}(0)|0\rangle\nonumber\\
&=&\sum_{m}v_{m}(\vec{x};X(T))\langle 0|\hat{b}_{m}(0)
T^{\star}\exp\{-\frac{i}{\hbar}\int_{0}^{T}
\hat{{\cal H}}_{eff}(t)dt\}\hat{b}^{\dagger}_{n}(0)|0\rangle
\nonumber\\
\end{eqnarray}
in agreement with (2.23).

In the adiabatic approximation,
where we assume the dominance of diagonal elements, we have 
(see also~\cite{kuratsuji}) 
\begin{eqnarray}
\psi_{n}(\vec{x},T; X(T))&\simeq&v_{n}(\vec{x};X(T))
\langle n|T^{\star}\exp\{-\frac{i}{\hbar}\int_{0}^{T}
\hat{{\cal H}}_{eff}(t)dt\}|n\rangle
\\
&\simeq&v_{n}(\vec{x};X(T))
\exp\{-\frac{i}{\hbar}\int_{0}^{T}[{\cal E}_{n}(X(t))
-\langle n|i\hbar\frac{\partial}{\partial t}|n\rangle]dt\}.
\nonumber
\end{eqnarray}
 
\section{Hidden local gauge symmetry}

All the results of the second quantization are
in principle reproduced by the first quantization in the present
single-particle problem by expanding the Schr\"{o}dinger 
amplitude $\psi(t,\vec{x})=\sum_{n}b_{n}(t)v_{n}(\vec{x},X(t))$ 
in terms of the instantaneous 
eigenfunctions in (2.10). One then analyzes simultaneous 
equations for the coefficients $\{b_{n}(t)\}$.
This equivalence is exemplified by the relation
(2.18). The possible advantages of the second quantized
formulation are thus mainly technical and conceptual ones, but 
we still obtain several interesting implications. First of all, 
the general 
geometric terms are explicitly and neatly formulated by the 
second quantization both for the path integral (2.12) and the 
operator formalism (2.18). 
Another technical advantage  is related to the phase freedom of
 the basis set in (2.10). The path integral formula (2.12) is 
based on the expansion
\begin{eqnarray}
\psi(t,\vec{x})=\sum_{n}b_{n}(t)v_{n}(\vec{x},X(t)),
\end{eqnarray}
and the starting path integral (2.2) depends only on the field
variable $\psi(t,\vec{x})$, not on  $\{ b_{n}(t)\}$
and $\{v_{n}(\vec{x},X(t))\}$ separately. This fact shows that 
our formulation contains an exact hidden local gauge symmetry 
\begin{eqnarray}
&&v_{n}(\vec{x},X(t))\rightarrow v^{\prime}_{n}(t; \vec{x},X(t))=
e^{i\alpha_{n}(t)}v_{n}(\vec{x},X(t)),\nonumber\\
&&b_{n}(t) \rightarrow b^{\prime}_{n}(t)=
e^{-i\alpha_{n}(t)}b_{n}(t), \ \ \ \ n=1,2,3,...,
\end{eqnarray}
where the gauge parameter $\alpha_{n}(t)$ is a general 
function of $t$. We tentatively call this symmetry 
"hidden local gauge symmetry" because it
appears due to the separation of the fundamental dynamical
variable $\psi(t,\vec{x})$ into two sets $\{ b_{n}(t)\}$
and $\{v_{n}(\vec{x},X(t))\}$.  One can confirm that the action 
\begin{eqnarray}
S=\int_{0}^{T}dt[&&
\sum_{n}b_{n}^{\star}(t)i\hbar\frac{\partial}{\partial t}
b_{n}(t)
+\sum_{n,m}b_{n}^{\star}(t)
\langle n|i\hbar\frac{\partial}{\partial t}|m\rangle
b_{m}(t)\nonumber\\
&&-\sum_{n}b_{n}^{\star}(t){\cal E}_{n}(X(t))b_{n}(t)]
\end{eqnarray}
and the path integral measure in (2.12) are both invariant under
 this gauge transformation. 
The Hamiltonian
\begin{eqnarray}
\hat{H}&=&\int \hat{\psi}^{\dagger}(t,\vec{x})
\hat{H}(\frac{\hbar}{i}\frac{\partial}{\partial\vec{x}}, 
\vec{x},  X(T))\hat{\psi}(t.\vec{x})d^{3}x dt\nonumber\\
&=&\sum_{n}{\cal E}_{n}(X(t))\hat{b}_{n}^{\dagger}(t)
\hat{b}_{n}(t)
\end{eqnarray}
is invariant under this local gauge symmetry, but the effective 
Hamiltonian (2.14) is not invariant under this transformation
\begin{eqnarray}
\hat{H}_{eff}(t) \rightarrow \hat{H}_{eff}(t)
+\sum_{n}\hbar \frac{\partial\alpha_{n}(t)}{\partial t}
\hat{b}_{n}^{\dagger}(t)\hat{b}_{n}(t).
\end{eqnarray}
This suggests that the conventional dynamical phase is 
manifestly gauge invariant and thus physical, whereas the 
geometric phase becomes physical after a non-trivial analysis
of gauge invariance. The above symmetry is exact as long as the 
basis set is not singular. In the present problem, the basis
set defined by (2.10) becomes singular on top of level 
crossing (see (4.4)), and thus the above symmetry is 
particularly useful in the general adiabatic approximation 
defined by the condition 
that the basis set (2.10) is well-defined. Of course, one may  
consider a new hidden local gauge symmetry when one defines a new
regular coordinate in the neighborhood of the singularity, and 
the freedom in the phase choice of the new basis set persists.

In our formulation, only $\{ b_{n}(t)\}$
are dynamical variables and thus it may be more natural to 
define the transformation by the second relation in (3.2), 
namely,
\begin{eqnarray}
b_{n}(t)=e^{i\alpha_{n}(t)}b^{\prime}_{n}(t), \ \ \ \ n=1,2,3,...
\end{eqnarray}
The field variable and the effective Hamiltonian are then 
transformed as
\begin{eqnarray}
\hat{\psi}(t,\vec{x})&=&\sum_{n}e^{i\alpha_{n}(t)}
\hat{b}^{\prime}_{n}(t)v_{n}(\vec{x},X(t))
=\sum_{n}
\hat{b}^{\prime}_{n}(t)v^{\prime}_{n}(t,\vec{x},X(t))
=\hat{\psi}^{\prime}(t,\vec{x}),
\nonumber\\
\hat{H}_{eff}(t)&=&\sum_{n}(\hat{b}^{\prime}_{n})^{\dagger}(t)
e^{-i\alpha_{n}(t)}
{\cal E}_{n}(X(t))e^{i\alpha_{n}(t)}\hat{b}^{\prime}_{n}(t)
\nonumber\\
&&-\sum_{n,m}(\hat{b}^{\prime}_{n})^{\dagger}(t)
e^{-i\alpha_{n}(t)}
\langle n|i\hbar\frac{\partial}{\partial t}|m\rangle
e^{i\alpha_{m}(t)}\hat{b}^{\prime}_{m}(t)\nonumber\\
&=&\sum_{n}(\hat{b}^{\prime}_{n})^{\dagger}(t)
{\cal E}_{n}(X(t))\hat{b}^{\prime}_{n}(t)
\nonumber\\
&&-\sum_{n,m}(\hat{b}^{\prime}_{n})^{\dagger}(t)
(\langle n|i\hbar\frac{\partial}{\partial t}|m\rangle)^{\prime}
\hat{b}^{\prime}_{m}(t) 
- \sum_{n}\hbar \frac{\partial\alpha_{n}(t)}{\partial t}
(\hat{b}^{\prime}_{n})^{\dagger}(t)
\hat{b}^{\prime}_{n}(t)
\end{eqnarray}
and the change of the basis set in (3.2) is realized. The action
(3.3) is confirmed to be form-invariant under the transformation
 (3.6). (In our practical applications below, the substitution
rules (3.2) give 
desired results without going through detailed manipulations.) 
Physically, this hidden gauge symmetry arises from the 
fact  that the 
choice of the basis set which specifies the coordinate in the 
functional space is arbitrary in field theory, as long as the 
coordinate is not singular. This local (i.e., time-dependent)
coordinate transformation, which is generally written in the form of (2.8), is thus  extended to an infinite dimensional 
unitary group $U(\infty)$. See also Ref.~\cite{giavarini}. The form of the action is 
generally changed under such a general transformation, though 
the physical contents of the theory are preserved. The possible 
subtlety under such a general unitary 
transformation can be analyzed by following the procedure 
in~\cite{fujikawa3}, but we 
do not expect  anomalous behavior in the present problem. In 
practical applications for  
generic eigenvalues $\{{\cal E}_{n}(X(t))\}$, the sub-group
\begin{eqnarray}
U=U(1)\times U(1)\times .....
\end{eqnarray}
as in (3.6) is useful, because it keeps the form of the action 
invariant and thus becomes a symmetry of quantized theory in the
 conventional sense. In particular, it is exactly 
preserved in the adiabatic approximation in which the mixing of
different energy eigenstates is assumed to be negligible and thus
the coordinates specified by (2.10) is always well-defined. 
 For a special case where the first eigenvalue
${\cal E}_{1}(X(t))$ has $n_{1}$-fold degeneracy, the second
eigenvalue ${\cal E}_{2}(X(t))$ has $n_{2}$-fold degeneracy, and 
so on, the sub-group
\begin{eqnarray}
U=U(n_{1})\times U(n_{2})\times ....,
\end{eqnarray}
which keeps the form of the action invariant, will be  useful.
See also Ref.~\cite{mostafazadeh} for a related analysis in the
framework of first quantization by using the notion of dynamical
invariants.
  
We emphasize once again that the above hidden local gauge 
symmetry (3.6) 
(or (3.2)) is an exact symmetry of quantum theory, and thus 
physical observables in the adiabatic approximation should 
respect this symmetry. Also, by using this local gauge freedom, 
one can choose the phase convention of the basis set 
$\{v_{n}(t,\vec{x},X(t))\}$ such that the analysis of geometric 
phases becomes most transparent.

Our basic observation is that $\psi_{n}(\vec{x},T; X(T))$ in 
the exact expression (2.23) (and also its adiabatic approximation
(2.29)) transforms under this hidden local gauge symmetry (3.6)
as
~\footnote{This shows that the state vector 
$\psi_{n}(\vec{x},t; X(t))$ stays in the same 
ray~\cite{streater} for an arbitrary hidden local gauge transformation.}
\begin{eqnarray}
\psi_{n}(\vec{x},T; X(T)) \rightarrow 
\psi^{\prime}_{n}(\vec{x},T; X(T))=e^{i\alpha_{n}(0)}
\psi_{n}(\vec{x},T; X(T))
\end{eqnarray}
{\em independently} of the value of $T$. 
This transformation is derived in (2.23) by using the 
representation (2.26)
\begin{eqnarray}
\psi^{\prime}_{n}(\vec{x},t; X(t))=
\langle 0|\hat{\psi}^{\prime}(t,\vec{x})
(\hat{b}^{\prime})^{\dagger}_{n}(0)|0\rangle=e^{i\alpha_{n}(0)}
\langle 0|\hat{\psi}(t,\vec{x})
\hat{b}^{\dagger}_{n}(0)|0\rangle
\end{eqnarray}
or by using 
\begin{eqnarray}
\psi^{\prime}_{n}(\vec{x},t; X(t))
=\int d^{3}y \langle \vec{x}
|T^{\star}\exp\{-\frac{i}{\hbar}\int_{0}^{t}
\hat{H}(\hat{\vec{p}}, \hat{\vec{x}},  
X(t))dt \}|\vec{y}\rangle\langle \vec{y}|n(0)^{\prime}\rangle.
\end{eqnarray}
with $v^{\prime}_{n}(0,\vec{y};X(0))
=\langle \vec{y}|n(0)^{\prime}\rangle
=e^{i\alpha_{n}(0)}v_{n}(\vec{y};X(0))$.
This transformation is also explicitly checked for the adiabatic
approximation (2.29). The transformation law (3.10) defined by
(3.11) or (3.12) is quite general since we assume that the set 
$\{v_{n}(\vec{y};X(0)) \}$ at $t=0$ is not singular.

Thus the product
\begin{eqnarray}
\psi_{n}(\vec{x},0; X(0))^{\star}\psi_{n}(\vec{x},T; X(T))
\end{eqnarray}
defines a manifestly gauge invariant quantity, namely, it is 
independent of the choice of the phase convention of the 
complete basis set $\{v_{n}(t,\vec{x},X(t))\}$. We employ this
(rather strong) gauge invariance condition as the basis of our analysis  of geometric phases.

Here, it may be appropriate to mention  briefly the difference 
between 
the present hidden local gauge symmetry and the freedom 
appearing in the analysis of the fiber bundles of  state
vectors in the Hilbert space. The states in quantum 
mechanics are represented by rays, namely, the states are 
specified up to {\em constant} phases~\cite{streater}. This may 
superficially appear to be a gauge symmetry. 
But the local time-dependent phases are not allowed in the ray 
space since the state multiplied by a time-dependent phase does 
not satisfy the Schr\"{o}dinger equation any more and thus goes 
outside the space of state vectors~\cite{samuel}. This differs 
from our hidden local gauge symmetry which is a symmetry 
of quantum theory and that the Schr\"{o}dinger amplitude 
$\psi_{n}(\vec{x},T; X(T))$ stays in 
the space of state vectors under an arbitrary hidden local gauge 
transformation of the basis set as is shown in (3.10). 
In the analysis of holonomy, it is common to consider a phase 
transformation of state vectors parametrized by 
$\alpha(X(t))$ in the precise adiabatic limit where the time 
dependence of $X(t)$ is negligible~\cite{simon}.
A further detailed comparison of the hidden local gauge 
symmetry to a gauge symmetry which appears in the 
definition of generalized geometric 
phases~\cite{aharonov, samuel} shall be given in Section 5.

For the adiabatic formula (2.29), the gauge invariant quantity 
(3.13) is given by
\begin{eqnarray}
&&\psi_{n}(\vec{x},0; X(0))^{\star}\psi_{n}(\vec{x},T; X(T))
\\
&&=v_{n}(0,\vec{x}; X(0))^{\star}v_{n}(T,\vec{x};X(T))
\exp\{-\frac{i}{\hbar}\int_{0}^{T}[{\cal E}_{n}(X(t))
-\langle n|i\hbar\frac{\partial}{\partial t}|n\rangle]dt\}
\nonumber
\end{eqnarray}
where we used the notation $v_{n}(t,\vec{x};X(t))$ to emphasize 
the use of arbitrary gauge in this expression.
We then observe that 
\begin{eqnarray}
v_{n}(0,\vec{x}; X(0))^{\star}v_{n}(T,\vec{x};X(T))
\exp\{-\frac{i}{\hbar}\int_{0}^{T}[
-\langle n|i\hbar\frac{\partial}{\partial t}|n\rangle]dt\}
\end{eqnarray}
is invariant under the hidden local gauge symmetry, and by choosing the gauge such that 
\begin{eqnarray}
v_{n}(T,\vec{x};X(T))=v_{n}(0,\vec{x}; X(0))
\end{eqnarray}
the prefacotor
$v_{n}(0,\vec{x}; X(0))^{\star}v_{n}(T,\vec{x};X(T))$ becomes real 
and positive. Note that we are assuming the cyclic motion of
the external parameter, $X(T)=X(0)$. (3.15) then becomes
\begin{eqnarray}
|v_{n}(0,\vec{x}; X(0))^{\star}v_{n}(T,\vec{x};X(T))|
\exp\{-\frac{i}{\hbar}\int_{0}^{T}[
-\langle n|i\hbar\frac{\partial}{\partial t}|n\rangle]dt\}
\end{eqnarray}
and the factor
\begin{eqnarray}
\exp\{-\frac{i}{\hbar}\int_{0}^{T}[{\cal E}_{n}(X(t))
-\langle n|i\hbar\frac{\partial}{\partial t}|n\rangle]dt\}
\end{eqnarray}
extracts all the information about the phase in (3.14) and 
defines a physical quantity. After this gauge fixing, the 
above quantity (3.17) is still invariant under residual gauge 
transformations satisfying the periodic boundary condition
\begin{eqnarray}
\alpha_{n}(0)=\alpha_{n}(T),
\end{eqnarray}
in particular, for a class of gauge transformations defined 
by $\alpha_{n}(X(t))$. Note that our gauge transformation in 
(3.6), which is defined by an arbitrary function 
$\alpha_{n}(t)$,  is more general. 

We here recognize an important 
difference between the conventional dynamical phase
\begin{eqnarray}
\exp\{-\frac{i}{\hbar}\int_{0}^{T}{\cal E}_{n}(X(t))dt\}
\end{eqnarray}
and the commonly defined geometric phase
\begin{eqnarray}
\exp\{-\frac{i}{\hbar}\int_{0}^{T}[-\langle n|i\hbar
\frac{\partial}{\partial t}|n\rangle]dt\}
\end{eqnarray}
though both of them are regarded as parts of the same total
Hamiltonian in the present formulation.
The conventional dynamical phase is manifestly gauge invariant,
whereas the conventional geometric phase is gauge covariant in
the sense that a gauge invariant meaning is assigned to it only 
for a specific choice of gauge, though the choice of the gauge 
is a sensible one.

For a {\em noncyclic} evolution but still adiabatic in the 
sense that the approximation (2.29) is valid,  the above gauge
invariant quantity (3.14)
\begin{eqnarray}
&&\psi_{n}(\vec{x},0; X(0))^{\star}\psi_{n}(\vec{x},T; X(T))
\\
&&=v_{n}(0,\vec{x}; X(0))^{\star}v_{n}(T,\vec{x};X(T))
\exp\{-\frac{i}{\hbar}\int_{0}^{T}[{\cal E}_{n}(X(t))
-\langle n|i\hbar\frac{\partial}{\partial t}|n\rangle]dt\}
\nonumber
\end{eqnarray}
still defines a physical quantity. But now $X(T)\neq X(0)$, and 
thus one cannot generally choose a gauge which makes 
$v_{n}(0,\vec{x}; X(0))^{\star}v_{n}(T,\vec{x};X(T))$ real and 
positive for all $\vec{x}$. Even in this case we can make 
\begin{eqnarray}
\int d^{3}x v_{n}(0,\vec{x}; X(0))^{\star}v_{n}(T,\vec{x};X(T))
\end{eqnarray}
real and positive by a suitable choice of the gauge 
$v_{n}(t,\vec{x};X(t)) \rightarrow 
v^{\prime}_{n}(t,\vec{x};X(t)) =
\exp[-i\alpha_{n}(t)]v_{n}(t,\vec{x};X(t))$. For such a choice 
of gauge, the factor 
\begin{eqnarray}
\exp\{-\frac{i}{\hbar}\int_{0}^{T}[{\cal E}_{n}(X(t))
-\langle n|i\hbar
\frac{\partial}{\partial t}|n\rangle^{\prime}]dt\}
\end{eqnarray}
extracts all the information about the phase of 
$\int d^{3}x \psi_{n}(\vec{x},0; X(0))^{\star}\psi_{n}(\vec{x},T;X(T))$,
and it gives an expression consistent with the Pancharatnam 
definition of geometric phase for a noncyclic 
evolution~\cite{samuel}. See Section 5 and also 
Refs.~\cite{pati2, garcia} for closely related analyses in the 
first quantization. The terminology "global gauge" was used in
Ref.~\cite{garcia} probably due to the $\vec{x}$ independence of
the gauge parameter. Our present formulation makes the origin of 
the local gauge invariance more transparent independently of 
approximation schemes. Note that our formula 
contains all the information about the phase factor and that the
 gauge invariant definition of the phase of 
 $ \int d^{3}x \psi_{n}(\vec{x},0; X(0))^{\star}\psi_{n}(\vec{x},T; X(T))$ as a line integral is unique up to 
gauge transformations with $\alpha_{n}(T)=\alpha_{n}(0)$.

We here recognize an important difference between the cyclic 
evolution and noncyclic evolution: The prefactor of the physical
 qauntity (3.14) for a cyclic evolution  can be made real and 
positive for arbitrary $\vec{x}$ by a suitable choice of hidden 
local gauge, whereas only the integrated prefactor (3.23) for a 
noncyclic evolution can be made real and positive by a choice of
 hidden local gauge. It is thus clear that the notion of 
geometric phase is of much more limited validity for a noncyclic
 evolution. 

For the most general noncyclic and non-adiabatic process,
the integrated gauge invariant  quantity is given by 
\begin{eqnarray}
&&\int d^{3}x \psi_{n}(\vec{x},0; X(0))^{\star}
\psi_{n}(\vec{x},T; X(T))
\nonumber\\
&&=\sum_{m}\int d^{3}x v_{n}(\vec{x};X(0))^{\star}
v_{m}(\vec{x};X(T))
\langle m|T^{\star}\exp\{-\frac{i}{\hbar}\int_{0}^{T}
\hat{{\cal H}}_{eff}(t)dt\}|n\rangle\nonumber\\
&&=\int d^{3}x v_{n}(\vec{x};X(0))^{\star}
\langle\vec{x}|T^{\star}\exp\{-\frac{i}{\hbar}\int_{0}^{T}
\hat{H}(\hat{\vec{p}}, \hat{\vec{x}},  
X(t))dt \}|n(0)\rangle
\end{eqnarray}
by using (2.23). Obviously this formula is exact but it is not clear if this general expression is useful in the practical analyses of geometric phases.
 
\section{Explicit example; two-level truncation}

It may be instructive to discuss a concrete example which shows 
how the hidden local symmetry works in the analysis of
 Berry's phases for noncyclic evolutions in general. We thus 
assume that the level crossing takes place only between 
the lowest two levels, and we consider the familiar idealized
model with only the lowest two levels. This simplification is 
expected to be valid in the neighborhood of the specific level 
crossing. 
The effective Hamiltonian to be analyzed 
in the path integral (2.6) is then defined  by the $2\times 2$
matrix $ h(X(t))=\left(E_{nm}(X(t))\right)$.
If one assumes that the level crossing takes place at the 
origin of the parameter space $X(t)=0$, one  analyzes
the matrix
\begin{eqnarray}
h(X(t)) = \left(E_{nm}(0)\right) + 
\left(\frac{\partial}{\partial X_{k}}E_{nm}(X)|_{X=0}\right)
X_{k}(t)
\end{eqnarray}
 for sufficiently small $(X_{1}(1),X_{2}(1), ... )$. By a time 
independent unitary transformation, which does not induce 
an extra geometric term, the first term is diagonalized.
In the present approximation, essentially the four dimensional 
sub-space of the parameter space is relevant, and after a 
suitable re-definition of the parameters by taking linear 
combinations of  $X_{k}(t)$, we write the matrix as~\cite{berry}
\begin{eqnarray}
h(X(t))
&=&\left(\begin{array}{cc}
            E(0)+y_{0}(t)&0\\
            0&E(0)+y_{0}(t)
            \end{array}\right)
        +g \sigma^{l}y_{l}(t)\nonumber\\
\end{eqnarray}
where $\sigma^{l}$ stands for the Pauli matrices, and $g$ is a 
suitable (positive) coupling constant. This parametrization in 
terms of the variables $y_{l}(t)$ is valid beyond the linear 
approximation, but the two-level approximation is expected to 
be valid only near the level crossing point.
 
The above matrix is diagonalized in the standard way as 
\begin{eqnarray} 
h(X(t))v_{\pm}(y)=(E(0)+y_{0}(t) \pm g r)v_{\pm}(y)
\end{eqnarray}
where $r=\sqrt{y^{2}_{1}+y^{2}_{2}+y^{2}_{3}}$  and
\begin{eqnarray}
v_{+}(y)=\left(\begin{array}{c}
            \cos\frac{\theta}{2}e^{-i\varphi}\\
            \sin\frac{\theta}{2}
            \end{array}\right), \ \ \ \ \ 
v_{-}(y)=\left(\begin{array}{c}
            \sin\frac{\theta}{2}e^{-i\varphi}\\
            -\cos\frac{\theta}{2}
            \end{array}\right)
\end{eqnarray}
by using the polar coordinates, 
$y_{1}=r\sin\theta\cos\varphi,\ y_{2}=r\sin\theta\sin\varphi,
\ y_{3}=r\cos\theta$. Note that our choice of the basis set 
satisfies
\begin{eqnarray}
v_{\pm}(y(0))=v_{\pm}(y(T))
\end{eqnarray}
if $y(0)=y(T)$ except for $(y_{1}, y_{2}, y_{3}) = (0,0,0)$, 
and $\theta=0\ {\rm or}\ \pi$; when one analyzes the behavior
near those singular points, due care needs to be exercised.
If one defines
\begin{eqnarray} 
v^{\dagger}_{m}(y)i\frac{\partial}{\partial t}v_{n}(y)
=A_{mn}^{k}(y)\dot{y}_{k}
\end{eqnarray}
where $m$ and $n$ run over $\pm$,
we have
\begin{eqnarray}
A_{++}^{k}(y)\dot{y}_{k}
&=&\frac{(1+\cos\theta)}{2}\dot{\varphi}
\nonumber\\
A_{+-}^{k}(y)\dot{y}_{k}
&=&\frac{\sin\theta}{2}\dot{\varphi}+\frac{i}{2}\dot{\theta}
=(A_{-+}^{k}(y)\dot{y}_{k})^{\star}
,\nonumber\\
A_{--}^{k}(y)\dot{y}_{k}
&=&\frac{1-\cos\theta}{2}\dot{\varphi}.
\end{eqnarray}
The effective Hamiltonian (2.14) is then given by 
\begin{eqnarray}
\hat{H}_{eff}(t)&=&(E(0)+y_{0}(t) + g r(t))\hat{b}^{\dagger}_{+}
\hat{b}_{+}
\nonumber\\
&+&(E(0)+y_{0}(t) - g r(t))\hat{b}^{\dagger}_{-}\hat{b}_{-}
 -\hbar \sum_{m,n}\hat{b}^{\dagger}_{m}A^{k}_{mn}(y)\dot{y}_{k}
\hat{b}_{n}
\end{eqnarray}
which is {\em exact} in the present two-level truncation.

In the conventional adiabatic approximation, one approximates
the effective Hamiltonian (4.8) by
\begin{eqnarray}
\hat{H}_{eff}(t)&\simeq& (E(0)+y_{0}(t) + g r(t))
\hat{b}^{\dagger}_{+}\hat{b}_{+}\nonumber\\
&&+(E(0)+y_{0}(t) - g r(t))\hat{b}^{\dagger}_{-}\hat{b}_{-}
\nonumber\\
&&-\hbar [\hat{b}^{\dagger}_{+}A^{k}_{++}(y)\dot{y}_{k}
\hat{b}_{+}
+\hat{b}^{\dagger}_{-}A^{k}_{--}(y)\dot{y}_{k}\hat{b}_{-}]
\end{eqnarray}
which is valid for 
\begin{eqnarray}
Tg r(t)\gg \hbar\pi,
\end{eqnarray}
where $\hbar\pi$ stands for the magnitude of the geometric term 
times $T$.
The Hamiltonian for $b_{-}$, for example, is then eliminated by 
a ``gauge transformation''
\begin{eqnarray}
b_{-}(t)=
\exp\{-(i/\hbar)\int_{0}^{t}dt[
E(0)+y_{0}(t) - g r(t) 
-\hbar A^{k}_{--}(y)\dot{y}_{k}] \} \tilde{b}_{-}(t)
\end{eqnarray}
in the path integral (2.12) with the above approximation (4.9), 
and the amplitude 
$\langle 0|\hat{\psi}(T)\hat{b}^{\dagger}_{-}(0)|0\rangle$, 
which corresponds to the probability amplitude in the first 
quantization, is given by (up to an eigenfunction 
$\phi_{E}(\vec{x})$ of 
$\hat{H}(\frac{\hbar}{i}\frac{\partial}{\partial\vec{x}},
\vec{x}, 0)$ in (2.3)) 
\begin{eqnarray}
\psi_{-}(T)&\equiv&\langle 0|\hat{\psi}(T)\hat{b}^{\dagger}_{-}(0)|0\rangle\nonumber\\
&=&\exp\{-\frac{i}{\hbar}\int_{0}^{T}dt[
E(0)+y_{0}(t) - g r(t) 
-\hbar A^{k}_{--}(y)\dot{y}_{k}] \}v_{-}(y(T))
\nonumber\\
&&\times 
\langle 0|\hat{\tilde{b}}_{-}(T)\hat{\tilde{b}}{}^{\dagger}_{-}(0)|0\rangle\nonumber\\
&=&\exp\{-\frac{i}{\hbar}\int_{0}^{T}dt[
E(0)+y_{0}(t) - g r(t) 
-\hbar A^{k}_{--}(y)\dot{y}_{k}] \}v_{-}(y(T))
\end{eqnarray}
with $\langle 0|\hat{\tilde{b}}_{-}(T)
\hat{\tilde{b}}{}^{\dagger}_{-}(0)
|0\rangle=\langle 0|\hat{\tilde{b}}_{-}(0)
\hat{\tilde{b}}{}^{\dagger}_{-}(0)
|0\rangle=1$. 
For a $2\pi$ rotation in $\varphi$ with fixed $\theta$, for 
example, the gauge invariant quantity (3.14) gives rise to  
\begin{eqnarray}
\psi_{-}(0)^{\star}\psi_{-}(T)&=&v_{-}(y(0))^{\star}v_{-}(y(T))
\nonumber\\
&&\times
\exp\{-\frac{i}{\hbar}\int_{0}^{T}dt[E(0)+y_{0}(t) - g r(t) 
-\hbar A^{k}_{--}(y)\dot{y}_{k}] \}\nonumber\\
&=&\exp\{i\pi(1-\cos\theta) \}
\exp\{-\frac{i}{\hbar}\int_{C_{1}(0\rightarrow T)}dt
[E(0)+y_{0}(t) - g r(t)] \}
\end{eqnarray}
by using (4.7) and $v_{-}(y(T))=v_{-}(y(0))$ in the present 
choice of gauge, and the path $C_{1}(0\rightarrow T)$
specifies the integration along the above specific closed path.
The first phase factor $\exp\{i\pi(1-\cos\theta) \}$
stands for the familiar Berry's phase~\cite{berry} and the 
second phase factor stands for the conventional dynamical 
phase.~\footnote{If
 the condition (4.10) is satisfied, the result  (4.13) is 
obtained  in a straightforward manner by  using the 
eigenfunctions in (4.4) consistently~\cite{fujikawa}. This 
expresses the fact that the choice of coordinates in the 
functional space does not matter in field theory and
one can use the most convenient coordinate. 
The analysis of the hidden local gauge symmetry in the present 
example will, however, 
help us compare the result in  the second quantized formulation 
to that in the first quantized formulation in which people are 
accustomed to the notion of holonomy and a very careful 
treatment of various phase factors. }
The phase factor (4.13)
is still invariant under a smaller set of gauge transformations
with
\begin{eqnarray}
\alpha_{-}(T)=\alpha_{-}(0)
\end{eqnarray}
and, in particular, for the gauge parameter of the form 
$\alpha_{-}(y(t))$~\cite{berry}. 
\normalsize
\begin{figure}[!htb]
 \begin{center}
    \includegraphics[width=10.9cm]{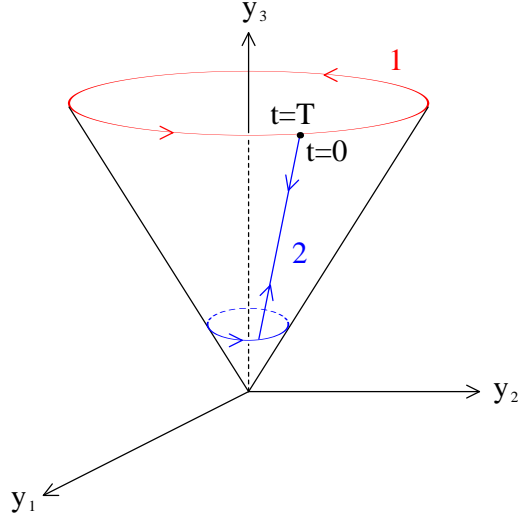} 
       \end{center}
\vspace{-9mm}      
 \caption{\small 
 The path 1 gives the conventional 
geometric phase as in (4.13) for a fixed finite $T$, 
whereas the path 2 gives a trivial geometric phase  as in (4.15)
 for a fixed finite $T$. Note that both of the paths cover the 
same solid angle $2\pi(1-\cos\theta)$.  }
\end{figure}

\vspace{1mm}

 In our previous papers~\cite{fujikawa, fujikawa2}, it has been
analyzed in detail how the conventional formula (4.13) is 
modified
if one deforms the contour in the parameter space for a fixed 
{\em finite} $T$. We here briefly comment on the main results. It was 
shown there that the amplitude in (4.13) is replaced by 
\begin{eqnarray}
\psi_{-}(0)^{\star}\psi_{-}(T)
&=&\exp\{-\frac{i}{\hbar}\int_{C_{2}(0\rightarrow T)}
dt[E(0)+y_{0}(t) - g r(t) 
-\hbar A^{k}_{--}(y)\dot{y}_{k}] \}\nonumber\\
&=&\exp\{-\frac{i}{\hbar}\int_{C_{2}(0\rightarrow T)}
dt[E(0)+y_{0}(t)-gr(t)] \}
\end{eqnarray}
by deforming the path 1 to the path 2 in the parameter space in 
Fig. 1. The path $C_{2}(0\rightarrow T)$ specifies the path 2 in 
Fig.1, and $v_{-}(y(T))=v_{-}(y(0))$ in the present choice of 
the gauge. Thus no geometric phase for the path $C_{2}$ for any 
fixed finite $T$. 
For $t=0$ or $t=T$, we start or end with the parameter 
region where the condition (4.10) for the adiabatic 
approximation is satisfied. But approaching
the infinitesimal neighborhood of the origin where the level 
crossing takes place, the condition is no more satisfied and
instead one has $Tgr\ll \hbar$.
In this region of the parameter space, $\hat{H}_{eff}$ 
in (4.8) or (4.9) is replaced by 
\begin{eqnarray}
\hat{H}_{eff}(t)&\simeq& (E(0)+y_{0}(t))
\hat{c}^{\dagger}_{+}\hat{c}_{+}\nonumber\\
&+&(E(0)+y_{0}(t))\hat{c}^{\dagger}_{-}\hat{c}_{-}
-\hbar\dot{\varphi} \hat{c}^{\dagger}_{+}\hat{c}_{+}
\end{eqnarray}
where one performed a unitary transformation
\begin{eqnarray}
\hat{b}_{m}=\sum_{n}U(\theta(t))_{mn}\hat{c}_{n}
\end{eqnarray}
with
\begin{eqnarray}
U(\theta(t))=\left(\begin{array}{cc}
            \cos\frac{\theta}{2}&-\sin\frac{\theta}{2}\\
            \sin\frac{\theta}{2}&\cos\frac{\theta}{2}
            \end{array}\right)
\end{eqnarray}
by assuming the validity of the two-level truncation in the 
infinitesimal neighborhood of the level crossing.
The diagonalization of the geometric terms in (4.16) corresponds
 to the use of eigenfunctions
\begin{eqnarray}
w_{+}=\left(\begin{array}{c}
            e^{-i\varphi}\\
            0
            \end{array}\right), \ \ \ \ \ 
w_{-}=\left(\begin{array}{c}
            0\\
            1
            \end{array}\right)
\end{eqnarray}
in the definition of geometric terms.
   
Based on this analysis, it was concluded 
in~\cite{fujikawa, fujikawa2} that the topological 
interpretation of the Berry's phase fails in the practical 
Born-Oppenheimer approximation where $T$ is identified with the 
period of the slower dynamical system. Also, the appearance of a seemingly non-integrable phase factor is consistent with the integrability of the Schr\"{o}dinger equation for a regular Hamiltonian.  
\\
\\
{\bf Geometric phase for noncyclic evolution}\\

We now analyze the geometric phase associated with a noncyclic
evolution on the basis of the explicit two-level truncation.
For the explicit example at hand, the starting gauge invariant 
formula (3.14) is given by
\begin{eqnarray}
&&\psi_{-}(0; X(0))^{\star}\psi_{-}(T; X(T))
\nonumber\\
&&=v_{-}(y(0))^{\star}v_{-}(y(T))
\exp\{-\frac{i}{\hbar}\int_{0}^{T}[E(0)+y_{0}(t) - g r(t)
-\hbar A^{k}_{--}(y)\dot{y}_{k}]dt\}
\nonumber\\
&&=v_{-}(y(0))^{\star}v_{-}(y(T))
\exp\{-\frac{i}{\hbar}\int_{0}^{T}
[-\hbar A^{k}_{--}(y)\dot{y}_{k}]dt\}\nonumber\\
&&\times \exp\{-\frac{i}{\hbar}\int_{0}^{T}[E(0)+y_{0}(t) 
- g r(t)]dt\}
\end{eqnarray}
by assuming the adiabatic approximation for the moment.
Here again, the combination
\begin{eqnarray}
v_{-}(y(0))^{\star}v_{-}(y(T))
\exp\{-\frac{i}{\hbar}\int_{0}^{T}
[-\hbar A^{k}_{--}(y)\dot{y}_{k}]dt\}
\end{eqnarray}
is invariant under hidden local gauge symmetry. For a noncycle
evolution, there is no simple choice of gauge which eliminates the factor $v_{-}(y(0))^{\star}v_{-}(y(T))$ altogether. 

In the present explicit example, we have by using (4.4)
\begin{eqnarray}
v_{-}(y(0))^{\star}v_{-}(y(T))
=\sin\frac{\theta(0)}{2}\sin\frac{\theta(T)}{2}
e^{i(\varphi(0)-\varphi(T))}+
\cos\frac{\theta(0)}{2}\cos\frac{\theta(T)}{2}.
\end{eqnarray}
By defining
\begin{eqnarray}
v_{-}(y(0))^{\star}v_{-}(y(T))=|v_{-}(y(0))^{\star}v_{-}(y(T))|
e^{i\Phi(T)}
\end{eqnarray}
one may perform a hidden local gauge transformation
\begin{eqnarray}
v_{-}(y(t)) \rightarrow v^{\prime}_{-}(t,y(t)) = 
e^{-i\alpha_{-}(t)}v_{-}(y(t))
\end{eqnarray}
such that 
\begin{eqnarray}
\alpha_{-}(T)-\alpha_{-}(0)=\Phi(T).
\end{eqnarray}
The net result is then the prefactor in (4.20) is replaced by 
$|v_{-}(y(0))^{\star}v_{-}(y(T))|$ and the geometric phase is 
shifted by $\Phi(T)$. Namely, we have
\begin{eqnarray}
&&\psi_{-}(0; X(0))^{\star}\psi_{-}(T; X(T))
\nonumber\\
&&=|v_{-}(y(0))^{\star}v_{-}(y(T))|\nonumber\\
&&\times
\exp\{-\frac{i}{\hbar}\int_{0}^{T}[E(0)+y_{0}(t)-gr(t)
-\hbar A^{\prime \ k}_{--}(y)\dot{y}_{k}]dt\}\nonumber\\
&&=|v_{-}(y(0))^{\star}v_{-}(y(T))|\nonumber\\
&&\times
\exp\{i\Phi(T)-\frac{i}{\hbar}\int_{0}^{T}[E(0)+y_{0}(t)-gr(t)
-\hbar A^{k}_{--}(y)\dot{y}_{k}]dt\}.
\end{eqnarray}
The definition of the geometric phase
\begin{eqnarray}
\int_{0}^{T} A^{\prime \ k}_{--}(y)\dot{y}_{k}dt
=\int_{0}^{T} A^{k}_{--}(y)\dot{y}_{k}dt + \Phi(T)
\end{eqnarray}
gives an expression consistent with the basic idea of 
Pancharatnam~\cite{samuel}. See Section 5 and also 
Refs.~\cite{pati2, garcia} for closely related analyses from
the different points of view.
Our formula contains all the information about the phase factor 
and that the gauge invariant definition of the phase of 
$\psi_{-}(0; X(0))^{\star}\psi_{-}(T; X(T))$ as a line integral 
is unique up to  gauge transformations with 
$\alpha_{-}(T)=\alpha_{-}(0)$.  

This quantity (4.26) is gauge invariant but path dependent in 
the parameter space for fixed 
$v_{-}(y(0))$, $v_{-}(y(T))$ and finite $T$. For example, for a path 
analogous to $C_{1}$ in Fig. 1 but now an open path (i.e., fixed
$\theta$ and $\varphi(T)-\varphi(0)<2\pi$), one has (see also
Ref.~\cite{garcia})
\begin{eqnarray}
\psi_{-}(0; X(0))^{\star}\psi_{-}(T; X(T))
&=&|v_{-}(y(0))^{\star}v_{-}(y(T))|\nonumber\\
&\times&
\exp\{i\Phi(T)+ i\frac{1}{2}(1-\cos\theta)(\varphi(T)-\varphi(0))
\}\nonumber\\
&\times&\exp\{-\frac{i}{\hbar}\int_{C_{1}}[E(0)+y_{0}(t) 
- g r(t)]dt\}.
\end{eqnarray} 
For a path analogous to $C_{2}$ in Fig.1 but now an open 
path (i.e., fixed $\theta$ and $\varphi(T)-\varphi(0)<2\pi$), 
one has
\begin{eqnarray}
\psi_{-}(0; X(0))^{\star}\psi_{-}(T; X(T))
&=&|v_{-}(y(0))^{\star}v_{-}(y(T))|\nonumber\\
&\times&
\exp\{i\Phi(T)+ i(\varphi(T)-\varphi(0))
\}\nonumber\\
&\times&\exp\{-\frac{i}{\hbar}\int_{C_{2}}[E(0)+y_{0}(t)-gr(t)]
dt\}.
\end{eqnarray} 
If one sets $\varphi(T)=\varphi(0)+2\pi$ and 
$v_{-}(y(T))=v_{-}(y(0))$ (and thus $\Phi(T)=0$), these formulas
 are reduced to the previous formulas for a cyclic evolution.

Since we analyzed the behavior in the neighborhood of the level 
crossing, the $\vec{x}$ dependent part is factored out as in 
(4.12) and the important difference between the cyclic and 
noncyclic evolutions noted in (3.23) does not explicitly appear 
in the present example.

\section{Comparison to gauge symmetry in the definition of 
generalized geometric phase}

We here explain a basic conceptual difference between the hidden
 local gauge symmetry and  a gauge symmetry (or equivalence 
class) which appears in the definition of generalized geometric 
phases by Aharonov and Anandan~\cite{aharonov} and also by 
Samuel and Bhandari~\cite{samuel}.

We first reformulate the treatments in these references from a
 view point of gauge symmetry by following
Refs.~\cite{aitchison, mukunda}.
The analysis in Ref.~\cite{aharonov} starts with the wave 
function satisfying 
\begin{eqnarray}
&&\int d^{3}x \psi(t,\vec{x})^{\star}\psi(t,\vec{x})=1,
\end{eqnarray}
and 
\begin{eqnarray}
&&\psi(T,\vec{x})=e^{i\phi}\psi(0,\vec{x})
\end{eqnarray}
with a real constant $\phi$. For simplicity we resrict our 
attention to the unitary time-development as in (5.1).
The condition (5.1) then implies the existence of a hermitian 
Hamiltonian
\begin{eqnarray}
i\hbar\frac{\partial}{\partial t}\psi(t,\vec{x})=
\hat{H}(t,\frac{\hbar}{i}\frac{\partial}{\partial\vec{x}},
\vec{x})\psi(t,\vec{x}).
\end{eqnarray}
The notion of generalized rays in refs.~\cite{aharonov, samuel} is based on the identification of all the vectors of the form 
\begin{eqnarray}
\{e^{i\alpha(t)}\psi(t,\vec{x})\}.
\end{eqnarray}
Note that they project 
$\psi(t,\vec{x})$ for each $t$, which means local in time
unlike the conventional notion of rays which is based on 
{\em constant} $\alpha$ ~\cite{streater}.
Since the conventional Schr\"{o}dinger equation is not invariant
under this equivalence class, we may consider an equivalence 
class of Hamiltonians
\begin{eqnarray}
\{ \hat{H} -\hbar\frac{\partial}{\partial t}\alpha(t)\}.
\end{eqnarray}
We next define an object 
\begin{eqnarray}
\Psi(t,\vec{x})\equiv\exp[i\int_{0}^{t}dt \int d^{3}x
\psi(t,\vec{x})^{\star}
i\frac{\partial}{\partial t}\psi(t,\vec{x}) ]
\psi(t,\vec{x})
\end{eqnarray}
which satisfies
\begin{eqnarray}
&&\Psi(0,\vec{x})=\psi(0,\vec{x}),\nonumber\\
&&\int d^{3}x \Psi(t,\vec{x})^{\star}
i\frac{\partial}{\partial t}\Psi(t,\vec{x})=0.
\end{eqnarray}
Under the equivalence class transformation (or gauge transfromation)
\begin{eqnarray}
\psi(t,\vec{x}) \rightarrow e^{i\alpha(t)}\psi(t,\vec{x}),
\end{eqnarray}
$\Psi(t,\vec{x})$ transforms as 
\begin{eqnarray}
\Psi(t,\vec{x}) \rightarrow e^{\alpha(0)}
\Psi(t,\vec{x}).
\end{eqnarray}
The quantity $\Psi(t,\vec{x})$ thus belongs to the same ray in 
the conventional sense under any gauge transformation. The 
properties (5.7) and (5.9) are valid independently of the precise
form of Schr\"{o}dinger equation (5.3), as we use only the 
property (5.1).

The gauge invariant quantity is then defined by
\begin{eqnarray}
\Psi(0,\vec{x})^{\star}\Psi(T,\vec{x})
=\psi(0,\vec{x})^{\star}\exp[i\int_{0}^{T}dt \int d^{3}x
\psi(t,\vec{x})^{\star}
i\frac{\partial}{\partial t}\psi(t,\vec{x}) ]
\psi(T,\vec{x})
\end{eqnarray}
by following our general prescription (3.13).
By a suitable gauge transformation 
\begin{eqnarray}
\psi(t,\vec{x})\rightarrow \tilde{\psi}(t,\vec{x})=
e^{-i\alpha(t)}\psi(t,\vec{x})
\end{eqnarray}
with
\begin{eqnarray}
\alpha(T)-\alpha(0)=\phi
\end{eqnarray}
we can make the prefactor in (5.10) 
\begin{eqnarray}
\psi(0,\vec{x})^{\star}\psi(T,\vec{x})\rightarrow
\tilde{\psi}(0,\vec{x})^{\star}\tilde{\psi}(T,\vec{x})
=e^{i\alpha(0)}\psi(0,\vec{x})^{\star}e^{-i\alpha(T)}
\psi(T,\vec{x})=|\psi(0,\vec{x})|^{2}
\end{eqnarray} 
real and positive for a cyclic evolution.
The above gauge invariant quantity is then given by 
\begin{eqnarray}
\Psi(0,\vec{x})^{\star}\Psi(T,\vec{x})
&=&|\psi(0,\vec{x})|^{2}\exp[i\int_{0}^{T}dt \int d^{3}x
\tilde{\psi}(t,\vec{x})^{\star}
i\frac{\partial}{\partial t}\tilde{\psi}(t,\vec{x}) ]
\end{eqnarray}
and the factor on the exponential extracts all the information
about the phase from the gauge invariant quantity.
This definition of the gauge invariant phase agrees
with the generalized geometric phase in~\cite{aharonov} by 
noting $\tilde{\psi}(0,\vec{x})=\tilde{\psi}(T,\vec{x})$. The 
phase factor in (5.14) is invariant under a residual gauge 
symmetry with 
\begin{eqnarray}
\alpha(T)=\alpha(0).
\end{eqnarray}
The phase factor in (5.14) is also written as 
\begin{eqnarray}
\beta=\oint dt \int d^{3}x
\tilde{\psi}(t,\vec{x})^{\star}
i\frac{\partial}{\partial t}\tilde{\psi}(t,\vec{x})
\end{eqnarray}
which makes the invariance under the gauge transformation 
(5.15) manifest,
but our basic formula (5.10) is invariant under a much larger 
class of gauge transformations. We can thus use the original
variable $\psi$ which satisfies (5.3) and write (5.10) as 
\begin{eqnarray}
\Psi(0,\vec{x})^{\star}\Psi(T,\vec{x})
=|\psi(0,\vec{x})|^{2}\exp[i\phi+\frac{i}{\hbar}
\int_{0}^{T}dt \int d^{3}x
\psi(t,\vec{x})^{\star}\hat{H}\psi(t,\vec{x})].
\end{eqnarray}
The phase on the exponential in (5.17) (and consequently, the 
phase in (5.14)) does not depend on the 
choice of the Hamiltonian in (5.5) since $\phi$ and 
the Hamiltonian are simultaneously changed by the parameter 
$\alpha(t)$.
The factor $(-1/\hbar)\int_{0}^{T}dt \int d^{3}x
\psi(t,\vec{x})^{\star}\hat{H}\psi(t,\vec{x})$ in (5.17) is
called a ``dynamical phase'' in~\cite{aharonov}. Note that the 
second relation in (5.7) does not play a major role in our 
formulation.
 
Next we comment on the generalized phase for a {\em noncyclic}
evolution~\cite{samuel}, namely, starting with (5.1) but the
relation (5.2) is modified to 
\begin{eqnarray}
\psi(T,\vec{x})\neq e^{\phi}\psi(0,\vec{x})
\end{eqnarray}
for any space-independent $\phi$. We can still consider the 
object (5.6) and the gauge invariant quantity (5.10)
\begin{eqnarray}
\Psi(0,\vec{x})^{\star}\Psi(T,\vec{x})
=\psi(0,\vec{x})^{\star}\exp[i\int_{0}^{T}dt \int d^{3}x
\psi(t,\vec{x})^{\star}
i\frac{\partial}{\partial t}\psi(t,\vec{x}) ]
\psi(T,\vec{x}).
\end{eqnarray}
The pre-factor $\psi(0,\vec{x})^{\star}\psi(T,\vec{x})$ now 
cannot be made real and positive for all $\vec{x}$ by any gauge 
transformation. One can still make the prefactor in the 
integrated quantity  
\begin{eqnarray}
\int d^{3}x\Psi(0,\vec{x})^{\star}\Psi(T,\vec{x})
=\int d^{3}x \psi(0,\vec{x})^{\star}\psi(T,\vec{x})
\exp[i\int_{0}^{T}dt \int d^{3}x
\psi(t,\vec{x})^{\star}
i\frac{\partial}{\partial t}\psi(t,\vec{x}) ]
\end{eqnarray}
real and positive. Namely, by defining 
\begin{eqnarray}
\int d^{3}x \psi(0,\vec{x})^{\star}\psi(T,\vec{x})=e^{i\Phi(T)}
|\int d^{3}x \psi(0,\vec{x})^{\star}\psi(T,\vec{x})|,
\end{eqnarray}
one may consider a gauge transformation
\begin{eqnarray}
\psi(t,\vec{x})\rightarrow \psi^{\prime}(t,\vec{x})
=e^{-i\alpha(t)}\psi(t,\vec{x})
\end{eqnarray}
such that
\begin{eqnarray}
\alpha(T)-\alpha(0)=\Phi(T).
\end{eqnarray}
One then obtains 
\begin{eqnarray}
&&\int d^{3}x\Psi(0,\vec{x})^{\star}\Psi(T,\vec{x})
\nonumber\\
&&=\int d^{3}x \psi^{\prime}(0,\vec{x})^{\star}\psi^{\prime}
(T,\vec{x})\exp[i\int_{0}^{T}dt \int d^{3}x
\psi^{\prime}(t,\vec{x})^{\star}
i\frac{\partial}{\partial t}\psi^{\prime}(t,\vec{x}) ]
\nonumber\\
&&=|\int d^{3}x \psi^{\prime}(0,\vec{x})^{\star}\psi^{\prime}
(T,\vec{x})|\exp[i\int_{0}^{T}dt \int d^{3}x
\psi^{\prime}(t,\vec{x})^{\star}
i\frac{\partial}{\partial t}\psi^{\prime}(t,\vec{x}) ]
\nonumber\\
&&=|\int d^{3}x \psi(0,\vec{x})^{\star}\psi(T,\vec{x})|
\exp[i\int_{0}^{T}dt \int d^{3}x
\psi^{\prime}(t,\vec{x})^{\star}
i\frac{\partial}{\partial t}\psi^{\prime}(t,\vec{x})].
\end{eqnarray}
The factor on the exponential extracts all the information 
about the phase factor of the integrated gauge invariant 
quantity. One can also use the variable $\psi$, which satisfies
the Schr\"{o}dinger equation (5.3), in the gauge invariant 
quantity (5.20) and write it as
\begin{eqnarray}
&&\int d^{3}x\Psi(0,\vec{x})^{\star}\Psi(T,\vec{x})
\nonumber\\
&&=|\int d^{3}x \psi(0,\vec{x})^{\star}\psi(T,\vec{x})|
\exp[i\Phi(T)+\frac{i}{\hbar}\int_{0}^{T}dt \int d^{3}x
\psi(t,\vec{x})^{\star}\hat{H}\psi(t,\vec{x})].
\end{eqnarray}
The phase factor in (5.25), which stands for the 
total phase increase of $\psi$ {\em minus} the ``dynamical 
phase'', agrees with the defining equation of the generalized 
phase in~\cite{samuel}.
Thus the phase on the exponential in (5.24) gives an alternative
 expression of the Pancharatnam phase difference as formulated 
in~\cite{samuel}. Note that the phase (5.24) is defined only for
 the integrated quantity as in (5.21), and it is 
invariant under the residual gauge symmetry satisfying (5.15).

Finally, we would like to compare the gauge symmetry appearing 
here to our  hidden local gauge symmetry in the case of  
adiabatic approximation (but with finite $T$), for which the 
correspondence becomes most visible. The basic correspondences 
are 
\begin{eqnarray}
\psi(t,\vec{x}) &\leftrightarrow& v_{n}(\vec{x};X(t)),
\end{eqnarray}
and 
\begin{eqnarray}
&&i\hbar\frac{\partial}{\partial t}\psi(t,\vec{x})=
\hat{H}(t,\frac{\hbar}{i}\frac{\partial}{\partial\vec{x}},
\vec{x})\psi(t,\vec{x})\nonumber\\
&& \leftrightarrow 
\hat{H}(\frac{\hbar}{i}\frac{\partial}{\partial\vec{x}}, 
\vec{x},X(t)) v_{n}(\vec{x};X(t))={\cal E}_{n}(t)
v_{n}(\vec{x};X(t))
\nonumber\\
\end{eqnarray}
with the equivalence class
\begin{eqnarray}
\{ e^{i\alpha(t)}\psi(t,\vec{x})\} &\leftrightarrow& 
\{ e^{i\alpha_{n}(t)}v_{n}(\vec{x};X(t))\}.
\end{eqnarray}
The quantity 
\begin{eqnarray}
&&\Psi(t,\vec{x}) = \exp[i\int_{0}^{t}dt \int d^{3}x
\psi(t,\vec{x})^{\star}
i\frac{\partial}{\partial t}\psi(t,\vec{x}) ]
\psi(t,\vec{x}) 
\end{eqnarray}
then corresponds to the quantity defined by 
\begin{eqnarray}
\Psi_{n}(\vec{x},t; X(t))&\equiv 
&\exp\{\frac{i}{\hbar}\int_{0}^{t}{\cal E}_{n}(X(t))dt\}
\psi_{n}(\vec{x},t; X(t))\nonumber\\
&=&
\exp[i\int_{0}^{t}
\langle n|i\frac{\partial}{\partial t}|n\rangle dt]
v_{n}(\vec{x};X(t)).
\end{eqnarray}
The physical observables in the cyclic evolution are then given
by, respectively,
\begin{eqnarray}
\Psi(0,\vec{x})^{\star}\Psi(T,\vec{x})
=\psi(0,\vec{x})^{\star}\psi(T,\vec{x})
\exp[i\int_{0}^{T}dt \int d^{3}x\psi(t,\vec{x})^{\star}
i\frac{\partial}{\partial t}\psi(t,\vec{x}) ]
\end{eqnarray}
and 
\begin{eqnarray}
&&\Psi_{n}(\vec{x},0; X(0))^{\star}\Psi_{n}(\vec{x},T; X(T))
\nonumber\\
&&=v_{n}(0,\vec{x}; X(0))^{\star}v_{n}(T,\vec{x};X(T))
\exp[i\int_{0}^{T}
\langle n|i\frac{\partial}{\partial t}|n\rangle dt]
\end{eqnarray}
where we used the notation $v_{n}(t,\vec{x};X(t))$ to emphasize 
the use of arbitrary gauge in the last expression.

The two formulations are thus very similar to each other, but 
there are several important differences. 
Most importantly, the true correspondence should be 
\begin{eqnarray}
\psi(t,\vec{x}) &\leftrightarrow& \psi_{n}(\vec{x},t; X(t))
\end{eqnarray}
instead of (5.26), since both of $\psi(t,\vec{x})$ and 
$\psi_{n}(\vec{x},t; X(t))$ in (2.23) stand for the 
Schr\"{o}dinger probability amplitudes. As a consequence of the difference between (5.26) and (5.33), a crucial difference in the conceptual level appears in the 
definition of equivalence class (or gauge symmetry). The hidden 
local gauge symmetry is consistent with the eigenvalue equation 
(5.27) and in fact it is an exact symmetry of quantized theory 
as was explained in Section 3. The Schr\"{o}dinger amplitude 
$\psi_{n}(\vec{x},t; X(t))$ is transformed under the hidden
local symmetry as 
\begin{eqnarray}
\psi_{n}(\vec{x},t; X(t)) \rightarrow 
\psi^{\prime}_{n}(\vec{x},t; X(t))=e^{i\alpha_{n}(0)}
\psi_{n}(\vec{x},t; X(t))
\end{eqnarray}
as is shown in (3.10). This hidden local symmetry is
 exactly preserved in the adiabatic approximation. 
(Although the Schr\"{o}dinger equation is satisfied only 
approximately in the adiabatic approximation,  it is a nature of
 the approximation.)  

In contrast, the equivalence class in the 
generalized definition (5.8)
changes the form of the Schr\"{o}dinger equation, and thus not a
symmetry of quantized theory in the conventional sense. In fact,
 the constant phase in $e^{i\alpha}\psi(t,\vec{x})$ does 
not change physics since it provides an overall constant phase 
for the state vector at all the times, but the time-dependent
 phase in $e^{i\alpha(t)}\psi(t,\vec{x})$  generally 
changes physics by providing different phases at different times
for the state vector. 
If one should take the equivalence class (5.4) literally, the 
{\em conventional} geometric phase would lose much of its significance  as is exemplified by the fact that the product of 
Schr\"{o}dinger wave functions
$\psi(0,\vec{x})^{\star}\psi(T,\vec{x})$ in (5.13) can be made 
real and positive by a suitable choice of gauge, though not all
is lost as the $\vec{x}$-dependence of $\psi(t,\vec{x})$ retains the information of the Hamiltonian.
By taking $\Psi(t,\vec{x})$ in (5.6) as a basic physical 
object, which is transformed by a constant phase under any 
gauge transformation, one can identify the generalized geometric
phases 
in~\cite{aharonov, samuel} by the consideration of gauge 
invariance alone, as we have explained  in this section by
following~\cite{aitchison, mukunda}.  The line integral 
along the ``vertical'' curve in~\cite{samuel} corresponds to
 the general gauge transformation (5.11) or (5.22), and the 
gauge symmetry which preserves the generalized geometric phases 
in~\cite{aharonov, samuel} corresponds to the residual gauge 
symmetry (5.15). These generalized geometric phases describe
 certain intrinsic properties of the class of Hamiltonians in 
(5.5) as is explained in detail in~\cite{aharonov, samuel}.
See also Refs.~\cite{aitchison, mukunda, pati1} for the further elaboration on these generalized geometric phases.

In comparison, the original analysis of holonomy by 
Simon~\cite{simon} is based on the gauge transformation
\begin{eqnarray}
\psi(t,\vec{x})\rightarrow e^{i\alpha(X(t))}\psi(t,\vec{x})
\end{eqnarray}
in the {\em precise} adiabatic limit (with 
$T \rightarrow \infty$) where the time-dependence of
$X(t)$ is negligible. In the precise adiabatic limit, it is 
known that the two formulations in (5.27) (when interpreted in
the sense of (5.35)) essentially 
coincide~\cite{aharonov} and thus two gauge symmetries with 
quite different origins give rise to the same result.

\section{Discussion} 

The analysis of geometric phases is reduced to the familiar
diagonalization of the Hamiltonian in the second quantized
formulation.   
The hidden local gauge symmetry, which is an exact symmetry of
quantum theory, becomes explicit in this formulation and we 
analyzed its full implications for 
cyclic and noncyclic evolutions in the study of geometric 
phases. We have shown that the general
prescription of Pancharatnam is consistent with the 
analysis on the basis of the hidden local gauge symmetry.
When one analyzes processes which are adiabatic only 
approximately, as in the practical Born-Oppenheimer 
approximation, the geometric phases cease to be purely 
geometrical. The notions of parallel transport and 
holonomy then become somewhat subtle, but our 
hidden local gauge symmetry is still exact and useful. The 
hidden local symmetry as formulated in this paper can thus 
provide a basic concept alternative to the notions of parallel 
transport and holonomy to analyze geometric phases associated 
with level crossing.
We have also explained a basic difference between the
 hidden local gauge symmetry and a gauge symmetry used in the 
definition of generalized geometric phases.

The notion of geometric phases is known to be exactly or 
approximately associated with a wide range of physical 
phenomena~\cite{shapere, review}. However, it is our opinion 
that the crucial differences of various physical phenomena,
 which are loosely associated with geometric phases, should be 
explicitly and precisely stated. The topological triviality of 
geometric phases associated with level crossing for any finite 
$T$, which is crucially different from the exact topological 
property of the Aharonov-Bohm phase, is one of those examples.
\\

I thank S. Deguchi for helpful discussions and A. Hosoya for
asking a connection of our formulation to that in 
Ref.~\cite{aharonov}. I also thank O. Bar for 
calling the non-level crossing theorem to my attention.

\end{document}